\documentclass[5p]{elsarticle}

 \usepackage{graphicx}
 \usepackage[font=footnotesize]{subfig} 
 \usepackage{relsize}

\usepackage{amssymb}
\usepackage{color}

\def\beq{\begin{equation}}
\def\eeq{\end{equation}}
\def\bea{\begin{eqnarray}}
\def\eea{\end{eqnarray}}

\newcommand*{\tabref}[1]{Table~\ref{tbl:#1}}
\newcommand*{\tablab}[1]{\label{tbl:#1}}
\newcommand*{\eqref}[1]{Eq.~(\ref{eq:#1})}
\newcommand*{\eqlab}[1]{\label{eq:#1}}
\newcommand*{\figref}[1]{Fig.~\ref{fig:#1}}
\newcommand*{\figlab}[1]{\label{fig:#1}}

\newcommand*{\secref}[1]{Section~\ref{sec:#1}}
\newcommand*{\seclab}[1]{\label{sec:#1}}

\def\VYP#1#2#3{{\bf #1}, #3 (#2)}  

\begin{document}

\begin{frontmatter}



\title{On the Radar detection of high-energy neutrino-induced cascades in ice; From Radar scattering cross-section to sensitivity}

\author[1]{Krijn D. de Vries}
\ead{krijndevries@gmail.com}
\author[1]{Paul Coppin}
\author[2]{Aongus \'O Murchadha}
\author[1,3]{Olaf Scholten}
\author[1]{Simona Toscano}
\author[1]{Nick van Eindhoven}
\address[1]{Vrije Universiteit Brussel, Dienst ELEM, IIHE, Pleinlaan 2, 1050 Brussels, Belgium}
\address[2]{Wisconsin  IceCube  Particle  Astrophysics  Center,  University  of  Wisconsin,  Madison,  WI 53706,USA}
\address[3]{University of Groningen, KVI-Center for Advanced Radiation Technology, 9747 AA Groningen, The Netherlands}

\begin{abstract}
In recent works we discussed the feasibility of the radar detection technique as a new method to probe high-energy cosmic-neutrino induced plasmas in ice. Using the different properties of the induced ionization plasma, an energy threshold of several PeV was derived for the over-dense scattering of a radio wave off the plasma. Next to this energy threshold the radar return power was determined for the different constituents of the plasma. It followed that the return signal should be detectable at a distance of several hundreds of meters to a few kilometers, depending on the plasma constituents and considered geometry. In this article we describe a more detailed modeling of the scattering process by expanding our model to include the full shower geometry, as well as the reflection off the under-dense plasma region. We include skin-effects, as well as the angular dependence of the scattered signal. As a first application of this more detailed modeling approach, we provide the effective area and sensitivity for a simplified detector setup. It follows that, depending on the detailed plasma properties, the radar detection technique provides a very promising method for the detection of neutrino induced particle cascades at energies above several PeV. Nevertheless, to determine the feasibility of the method more detailed information about the plasma properties, especially its lifetime and the free charge collision rate, are needed.
\end{abstract}

\end{frontmatter}

\section{Introduction}
In 2013 the IceCube neutrino observatory for the first time detected high-energy cosmic neutrinos~\cite{I3_2013sc}. At the highest energies these cosmic neutrinos become very rare, and above several PeV IceCube runs out of statistics. For the detection of events at energies larger than several PeV, an even larger effective volume than the cubic kilometer instrumented by IceCube is needed. It follows that due to its long attenuation length, the radio signal is an excellent probe for cosmic-neutrino-induced cascades at the highest energies. Nevertheless, the currently existing Askaryan radio detectors such as ARA~\cite{ARA} and ARIANNA~\cite{ARIANNA} only start to become sensitive at several EeV where the GZK-flux is expected~\cite{Grei66,Zat66}. 

In this work we discuss the radar reflection technique as a possible method to fill the currently existing energy gap between several PeV and a few EeV. The radar detection method for cosmic-ray particle cascades in air was first proposed in the 1940's~\cite{Bla40} and has been revised at the beginning of this century~\cite{Bar93,Gor01}. Even-though several radar detection experiments have been conducted ever since~\cite{Sug68,Vin01,Lyo03,Ter09,Tak10,Oth11,Bel13,Abb16} along with several new modeling attempts~\cite{Bak10,Tak11,Sta13}, no conclusive evidence for the detection of a cosmic-ray air shower has been found up to this moment. Recently it was pointed out that due to the large free charge collision rate in air the scattering efficiency is expected to drop significantly, and as such the scatter becomes too weak to observe~\cite{Sta15}. More recently suggestions were done to use the radar technique for the detection of particle cascades in dense media such as ice and rock~\cite{Chi13,dVries15}. For these media, the free charge collision rates and lifetime of the free charge plasma are currently unknown, and have to be determined experimentally.

In previous work~\cite{dVries15}, we considered the feasibility of the radar detection technique in ice. When a high-energy cosmic neutrino interacts in a natural ice-sheet, a high-energy particle cascade will be induced. While propagating, the cascade will loose most of its energy by ionizing the medium. Based on measurements performed in the 1980's~\cite{Ver78,Kun80,Haa83a,Haa83b}, we considered two different constituents of the induced ionization plasma. Next to a rather short-lived electron plasma, a long-lived proton plasma was considered. Using the obtained lifetimes of the plasma, we were able to derive an energy threshold for the over-dense scattering off the plasma of 4~PeV for the electron plasma and 20~PeV for the proton plasma. Next to the energy threshold, we determined the radar return power as a function of distance to the cascade. It was seen that depending on the geometry and constituent of the plasma, the return signal should be detectable at distances between several hundreds of meters up to several kilometers, making the radar detection method a very promising technique for the detection of high-energy cosmic neutrinos above several PeV. 

The calculations done in our previous work were performed considering a simplified cascade geometry. Furthermore, we only considered the scattering off the over-dense ionization plasma. In this article, we extend our model to realistic cascade geometries and include the reflection due to the under-dense scattering. A more detailed scattering model is considered accounting for skin-effects, the angular dependence of the scattering, and the lifetime of the plasma~\cite{dVries17a,dVries17b}. This allows us to calculate the radar scattering cross-section without having to assume an empirically derived thin wire approximation~\cite{Gor01,Cri65}. Furthermore, we discuss the scattering off the high-energy shower particles and argue that this contribution is small with respect to the over-dense scatter, and hence can be neglected.

As a first application of the improved calculation, we determine the effective area and sensitivity for a simplified detector set-up. We show that the radar detection technique starts to become sensitive at PeV energies with increasing sensitivity toward higher energies. Nevertheless, the sensitivity will strongly depend on the plasma properties such as its lifetime and the free charge collision rate which are unknown. Therefore, to confirm these results more detailed information is needed about the free charge plasma. To determine these properties a beam-test experiment has been conducted at the Telescope Array (TA) Electron Light Source (ELS) facility~\cite{Shi09,dVries15b}, searching for a radar scatter off the induced ionization plasma which remained after a block of ice was irradiated by a high-energy electron beam (O(PeV) equivalent energy). During these experiments a first hint for a scattered signal has been observed~\cite{dVries15b,dVries17c}. To confirm such a scatter and determine the plasma properties in more detail additional experimental efforts are needed. Currently, such an experimental effort is scheduled to take place at the Stanford Linear Accelerator Center early 2018~\cite{Pro18a,Pro18b}.

\section{The plasma}
A high-energy cosmic neutrino can induce a particle cascade in ice. In our modeling we will consider the full three-dimensional particle density profile. For the longitudinal development of the cascade, we will use the NKG-parameterization to model the total number of particles in the cascade, given as function of the penetration depth $X\mathrm{(g/cm^2)}$~\cite{Kam58,Grei65},
\beq
N(X)=\frac{0.31\;\exp[(X/X_{0})(1-1.5\ln s)]}{\sqrt{\ln(E/E_{crit})}}.
\eqlab{NKG}
\eeq 
Here $X_0=36.08\;\mathrm{g/cm^2}$ is the electron radiation length in ice, $E_{crit}=0.0786\;\mathrm{GeV}$ the critical energy for electrons in ice, and $s$ denotes the shower age given by,
\beq 
s(X)=\frac{3X/X_0}{(X/X_0)+2\ln(E/E_{crit})}.
\eqlab{age}
\eeq
For the lateral particle distribution, we consider a radial symmetry $w(r)=2\pi w(\vec{r})$ and follow the parameterization given in~\cite{Wer12} which is used for cosmic-ray air showers,
\beq
w(r)=\frac{\Gamma(4.5-s)}{\Gamma(s)\Gamma(4.5-2s)}\left(\frac{r}{r_0}\right)^{s-1}\left(\frac{r}{r_0}+1\right)^{s-4.5}.
\eeq
Here $r\;\mathrm{(cm)}$ is the radial distance with respect to the shower axis, and $r_0$ is the moli\`ere radius. Converting the in air distance to depth (see appendix B of~\cite{dVries15}), the in-ice value is approximated by $r_0\approx 7$~cm. So-far, we only discussed the high-energy particle cascade. Nevertheless, we are mostly interested in the ionization plasmas induced by this high-energy particle cascade. To determine the ionization plasma, we have to consider the energy loss due to ionization by a high-energy particle inside the cascade. Throughout this article we will use the typical value of $E_{ion}=2~\mathrm{MeV/g/cm^2}$ which is given for a minimum ionizing particle. The total energy loss due to ionization is now given by,
\beq
E_{loss}^{ion}=\int N(X)E_{ion}\;dX.
\eeq
Performing the integral over a wide range of cascade energies shows that typically $90\%$ of the primary shower energy is lost to ionization. This value agrees very well with results from more detailed Monte-Carlo simulations. 

In general the low-energy ionization plasma will diffuse into the medium. The effect of this diffusion will mostly depend on the lifetime of the plasma and in general be rather small. This is mainly due to the fact that the plasma will consist of low energetic particles with energies around the ionization energy of O(10)~eV. The maximum diffusion of the plasma can be estimated by,
\beq
r_{\rm diff}^{\perp}=\left(\frac{m_e}{m_p} \frac{E}{E_0}\right)^{1/2} v_e\sin(\theta)\tau_p,
\eeq
where $v_e\approx 6\cdot10^{-2}~\mathrm{cm/ns}$ is the speed of a free electron with an energy of $E_0=1$~eV, $E$ is the energy in electron volts, $m_p$ denotes the effective mass of the plasma constituent, $\theta$ is the angle of the plasma particle with respect to the direction of the cascade, and $\tau_p$ is the plasma lifetime in nanoseconds. Assuming a collision-less plasma where the free charge will have its momentum in the direction of the ionizing particle, $\sin{(\theta)}$ can be approximated by, $\sin(\theta)\approx r_0/L\approx 10^{-2}$, the ratio between the moli\`ere radius and the length of the cascade. As will be outlined in the following section, in this article we consider two different ionization plasmas. First there is a free electron plasma for which a lifetime up to several tens of nanoseconds is considered. In addition, we consider a free plasma with its properties equal to that of free protons, for which a longer lifetime up to 1~$\mathrm{\mu s}$ will be used. Taking a typical energy of $E=10$~eV for the free electrons and a lifetime of 20~ns, the diffusion is estimated to be of the order of $r_{\rm diff}^{\perp}=O(10^{-2})$~cm. For the proton plasma, using a longer lifetime of $1$~$\mu$s, a similar radial diffusion distance is obtained. Therefore, in the following we will neglect diffusion effects and consider a moli\`ere radius of $r_0=7$~cm for the ionization plasma.

\subsection{The over-dense plasma}
When determining the radar scattering cross-section for the scattering off a plasma, we need to consider two different regimes. The so-called over-dense regime and the under-dense regime. The over-dense region is defined by the condition,
\beq
\eqlab{eq_wp}
\omega_p > \omega_d,
\eeq
where $\omega_p$ denotes the plasma frequency, and $\omega_d$ gives the observation frequency. The plasma frequency is given by~\cite{Hecht},
\beq
\omega_p=8980\sqrt{\frac{m_e}{m_p}n_e},
\eqlab{op}
\eeq
where $m_e$ is the electron mass, $m_p$ the (effective) mass of the plasma constituent, and $n_e(\mathrm{cm^{-3}})$ denotes the free charge density of the plasma. In case of over-dense scattering, the incoming wave will scatter off the surface of the plasma volume. In case of under-dense scattering defined by the condition that $\omega_p < \omega_d$, the incoming wave will scatter off the individual electrons leading to a decreased radar scattering cross-section.
As will be pointed out in detail in~\secref{lt}, the detection frequency is limited by the lifetime, $\tau_p$, of the plasma cloud. The condition for over-dense scattering in~\eqref{eq_wp} therefore becomes,
\bea
\eqlab{eq_el}
\omega_p > \omega_d > 1/\tau_p.
\eea
It should be noted that the electron density, and hence the plasma frequency, scales with the energy of the primary cascade inducing particle. The condition given in~\eqref{eq_el}, therefore immediately leads to an energy threshold for the radar reflection off the over-dense plasma. In~\cite{dVries15}, this energy threshold was approximated assuming an isotropic distribution within the inner two centimeters of the particle cascade containing approximately half of the total number of particles in the plasma. 

The electron plasma was found to be limited by its lifetime, where a rather conservative value of $\tau_e=1$~ns was used following~\cite{Ver78,Haa83a}. In~\cite{Kun80,Haa83b}, next to the electron plasma a long lived plasma was found consistent with free protons. For the free proton plasma, due to its long lifetime, the limiting factor used in~\cite{dVries15}, was given by the size of the plasma cloud of approximately 5~m. As will be shown in~\secref{dif}, the dimension of the cascade determines the angular distribution of the scattered signal and is not a measure for the scattering to actually occur, and hence the energy threshold for the free proton plasma was overestimated. Using the obtained lifetimes and plasma dimensions, in~\cite{dVries15} an energy threshold was obtained of 4~PeV for the electron plasma, and 20~PeV for the proton plasma.

In this article, we use the full three-dimensional cascade profile (assuming radial symmetry) to determine the radar return power and the size of the over-dense regime. In~\figref{profile1}, we plot the ionization plasma for the situations sketched above. The electron plasma is shown for a 4~PeV cascade inducing particle observed at 1~GHz, where the proton plasma is given for a 20~PeV cascade inducing particle detected at 50~MHz. As expected from the approximations done in~\cite{dVries15}, a significant part of the plasma is over-dense in both situations.
\begin{figure*}[t]
\subfloat[]{%
  \includegraphics[width=.45\linewidth]{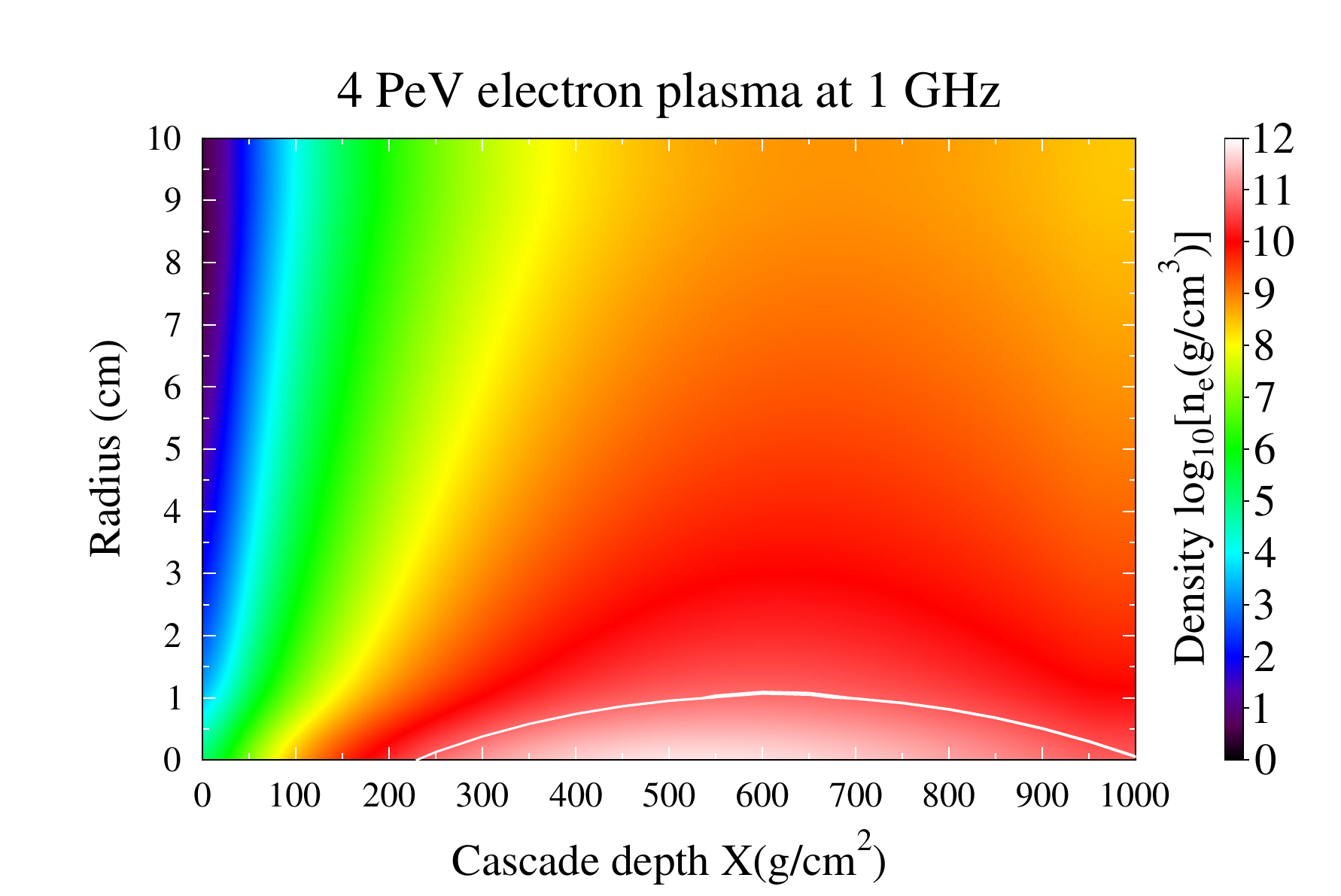}%
}\hfill
\subfloat[]{%
  \includegraphics[width=.45\linewidth]{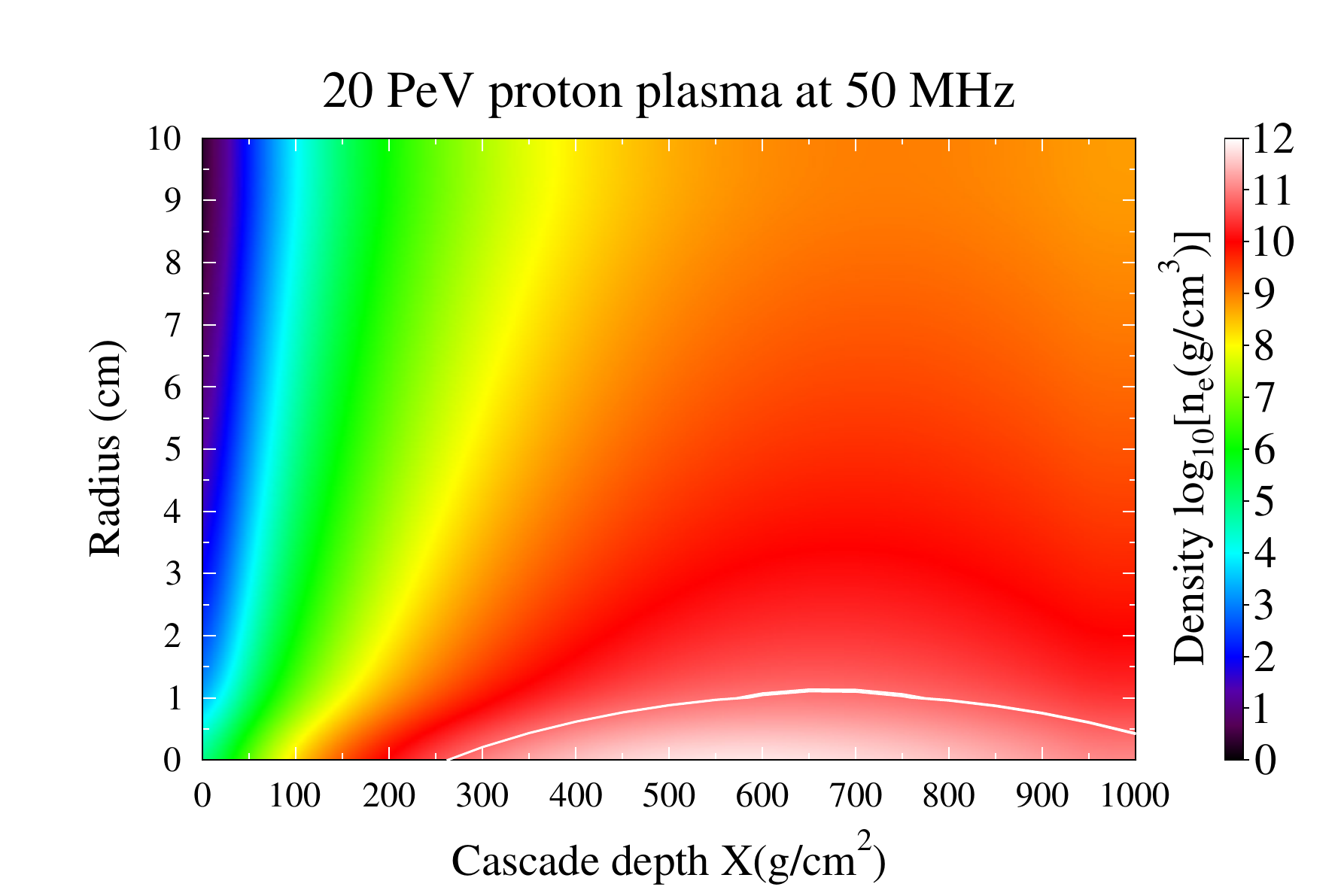}%
}
\caption{(a)~The electron plasma modeled for a 4~PeV cascade inducing particle. (b)~The proton plasma modeled for a 20~PeV cascade inducing particle (figure taken from~\cite{dVries17a}). The over-dense plasma defined by~\eqref{eq_wp}, is bounded by the white line.}
\figlab{profile1}
\end{figure*}
\section{Radio wave scattering}
In~\cite{dVries15}, we only considered the scattering of a radio wave off the over-dense part of the plasma. For the scattering we assumed a simplified cylindrical geometry, with an isotropic particle density. Furthermore, we assumed isotropic scattering and skin effects were ignored. In this section we present a more realistic approach, taking into account the scattering off the under-dense plasma as well as a more realistic approach to model the over-dense scattering.

\subsection{Under-dense scattering}
In the previous section we modeled the plasma, considering both the over-dense part as well as the under-dense part of the plasma. This allows us to numerically evaluate the total number of free charges in the under-dense regime. For the under-dense part of the plasma, the free charge density is too low for the individual charges to affect each other. Therefore, the scattering can be seen as a superposition of the scattering off the individual charges in the plasma. The radar cross-section for this scattering is given by the Thomson cross-section describing the scattering of an electromagnetic wave on a free charge of mass $m_p$ with respect to the electron mass $m_e$. The Thomson scattering cross-section is given by~\cite{thomson},
\beq
\sigma_{\rm T}=\left(\frac{m_e}{m_p}\right)^2 0.665\cdot10^{-28}\;\mathrm{m^2}.
\eeq
For the total radar scattering cross-section we have to take into account the phase lag between the reflected signals of the individual electrons. This gives an effective cross-section equal to
\beq
\sigma_{\rm ud}=N^{\alpha}\sigma_{\rm T},
\eqlab{sud}
\eeq
where $N$ denotes the total number of free charges in the under-dense plasma, and the power $\alpha$ follows from the phase delay between the individual particles, which is taken into account in the simulation. Following this procedure, the power $\alpha$ depends on the detection frequency, as well as the plasma lifetime and ranges between $\alpha=1$ for completely incoherent scattering and $\alpha=2$ for coherent scattering.


\subsection{Over-dense scattering}
\seclab{dif}
Now that we obtained the total under-dense radar scattering cross-section, we will consider the cross-section for the over-dense case. In~\cite{dVries15}, we considered an isotropic particle distribution within a cylindrical tube containing approximately $50\%$ of the total number of particles. Consequently for the radar scattering cross-section we considered a thin-wire approximation obtained from~\cite{Gor01,Cri65}, where skin-effects were ignored. In this section we detail the procedure presented in~\cite{dVries17a}, and reconstruct the radar scattering cross-section by considering the projected area of the plasma directly including skin effects. This will be done by considering regions of equal density which will be approximated by a cylinder of length $L$ with radius $r$.

For the determination of the over-dense radar scattering cross-section, we have to convolve the cylinder area with a geometrical factor taking into account the angle of the incoming wave with respect to normal incidence, as well as the polarization angle in the cascade plane. Furthermore, we can consider three different regimes for the scattered signal. In case the wavelength of the incoming wave is large compared to the dimension of the plasma, the signal will be scattered isotropically. In case the wavelength is much smaller than the dimension of the plasma, the plasma will act like a perfect mirror. In our situation most of the time the wavelength is of the order of the size of the plasma cloud and a 'slit-like' interference can be expected. Therefore an interference factor $f_{\rm dif}$ is introduced which gives the relative intensity with respect to isotropic scattering. These effects will be incorporated in the geometry factor $f_{\rm geom}$, which is given by,
\beq
f_{\rm geom}=(1-\vec{e}_{tc}\cdot\vec{e}_c)(\vec{e_t}\cdot\vec{e_c})f_{\rm dif},
\eeq
where $\vec{e}_{tc}$ denotes the unit vector pointing from the transmitter to the cascade, $\vec{e}_c$ denotes the cascade direction (polarization), and $\vec{e}_t$ denotes the polarization of the transmitted signal in the cascade plane. Since the longitudinal dimension of the plasma is in general large compared to the radial dimension, and the observer will be positioned in the far-field ($R >> \lambda$), the interference factor $f_{\rm dif}$ can be approximated by Fraunhofer diffraction from a single slit~\cite{Hecht},
\beq
f_{\rm dif}=\frac{I(\alpha)}{<I(\alpha)>},
\eeq
using,
\beq
I(\alpha)= \mathrm{sinc}^2(\beta).
\eeq
Here $\beta=(\pi L / \lambda)\sin{\alpha}$, where $L$ denotes the length of the over-dense plasma region, and $\alpha$ is the angle relative to the angle of total internal reflection in the cascade plane (see~\figref{diffraction}).
\begin{figure}
  \centering{
  \includegraphics[width=.45\textwidth]{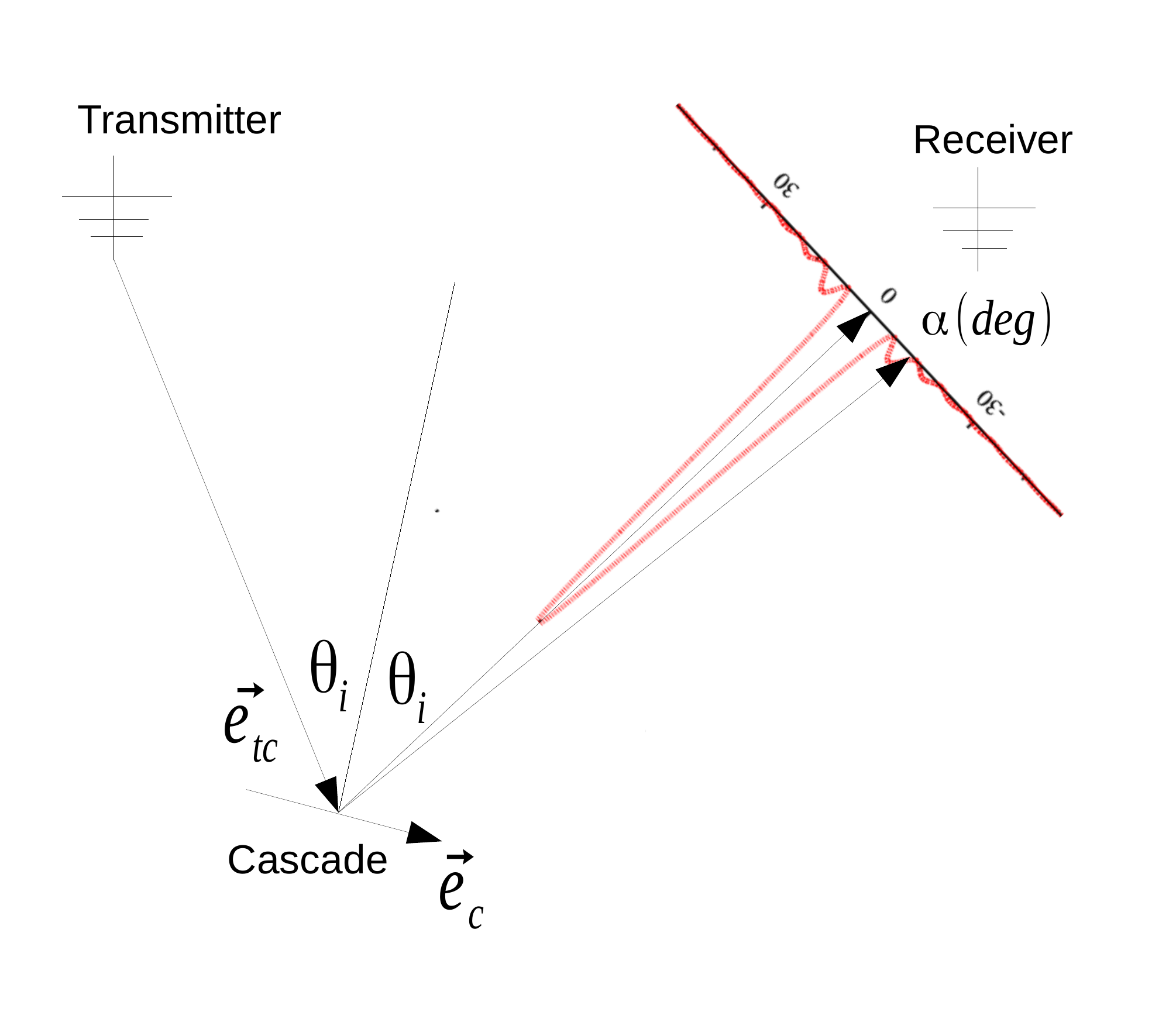}}
\caption{The scatter geometry. The transmitted signal travels toward the cascade, after which it gets reflected to the receiver located at an angle $\alpha$ from the angle of total internal reflection. The red (dashed) line indicates the relative intensity in linear scale as function of $\alpha$.}
\figlab{diffraction}
\end{figure}

Furthermore, we will directly incorporate skin effects into the calculation of the over-dense radar scattering cross-section. Considering $N$ regions with approximately constant density, the total over-dense radar scattering cross-section can be written as,
\beq
\sigma_{\rm od}=\sum\limits_{i=1}^{N} A_{\rm c}^i\times f_{\rm geom} \times f_{\rm skin}^i
\eeq
Where $A_c^i$ is the area of the i-th cylinder given by,
\beq
A_{\rm c}^{i}=\frac{1}{2}(2\pi) r_{\rm mx}^{i} L^i,
\eeq
with $r_{\rm mx}^{i}$ the maximum radius of the over-dense plasma and $L^i$ denotes the length of the over-dense plasma determined by either the length of the cascade or the lifetime of the plasma. An additional factor $0.5$ is included since in reality our region of constant density is not cylindrical, but cigar shaped as can be seen from~\figref{profile1}. 
To include skin effects, we have to consider the power of an electromagnetic wave impinging an over-dense plasma. This power will drop as a function of the distance $x$ inside the plasma following,
\beq
P(x)=P_0e^{-x/\delta},
\eeq
where $\delta=c/2w_p$ is the skin depth of the plasma. Hence,
\beq
P_{\rm sc}(x)=P_0(1-e^{-x/\delta}),
\eqlab{Pabs}
\eeq
is the absorbed power exiting the plasma. This power will be re-emitted by the plasma, which is the considered over-dense scattering. It follows immediately that if the dimension of the plasma is small with respect to the plasma wavelength $\lambda_p=w_p/c$, that only part of the incoming power will be scattered. Due to the non-constant density of the plasma, the skin depth, $\delta(x)$, will be a function of $x$. The scattered fraction is therefore calculated by defining layers of width $\Delta x$, for which the particle density and hence the skin depth $\delta$, becomes independent of $x$. The fraction $f^{i}$ of the power reflected in the (i)-th layer is given by the fraction of power that remains after crossing the previous layers, $f_{\rm rem}^i=(1-(\mathsmaller{\sum\limits_{j=0}^{i-1}}f^{j}))$, convolved with the amount of power absorbed in this layer of thickness $\Delta x$,
\beq
f^{i}_{\rm skin}=f_{\rm ref}^i(1-e^{-\Delta x/\delta^i}).
\eeq
The total over-dense radar scattering cross-section can now be expressed as,
\bea
\sigma_{\rm od}&=&\sum\limits_{i=1}^{N} A_{c}^i\times f_{\rm geom} \times f_{\rm skin}^i\\
&=&\sum\limits_{i=1}^{N}\mathlarger{\mathlarger{\{}}\pi r_{\rm mx}^{i} L^i(1-\vec{e}_{tc}\cdot\vec{e}_c)(\vec{e_t}\cdot\vec{e_c})f_{\rm dif}\nonumber\\
&&\;\;\;\;\;\;(f_{\rm ref}^i)(1-e^{-\Delta x/\delta^i})\mathlarger{\mathlarger{\}}.}
\eqlab{sodt}
\eea



\subsection{The high-energy shower front}
So far we only considered scattering off the static ionization plasma. Next to the ionization plasma the radio wave will also scatter off the high-energy particles in the cascade itself. To estimate the strength of this scatter, we first have to consider the charge density. In~\cite{dVries15}, we show that the charge density of the high-energy particles in the cascade front is a factor $10^5$ below that of the ionization plasma. Therefore, the scatter will be highly suppressed. Depending on the observation angle with respect to the cascade a boosting of the emission can be expected. To estimate the boosting effect, one has to consider that along with the boosting, the frequency of the return signal is also shifted, and the emission will be beamed. For a frequency boost $\nu \rightarrow \nu'=f_b\cdot\nu$, the emitted power is boosted with a factor $f_{\rm b}^2$. For the boosting to become efficient compared to the scattering off the over-dense ionization plasma, a factor $f_b^{eff}$ equal to the ratio of the effective area of the over dense ionization plasma $A_{\rm od}=O(1 - 10^{-3}~{\rm m}^2)$ and the under-dense shower front particles is needed,
\beq
f_b^{eff}\approx \sqrt{\frac{A_{ \rm od}}{N_{ \rm cas}^\alpha \sigma_{\rm T}}}.
\eeq
For cascade energies in the PeV-EeV region, the total number of particles in the high-energy cascade front equals $N_{\rm cas}\approx 10^6 - 10^9$ respectively, leading to required boost factors $f_b^{eff}>10^4$. At the corresponding boosted frequencies, the ice becomes opaque. Furthermore, the opening angle of this emission will be extremely small such that a very dense instrumentation over the full solid angle would be needed to observe the radar scatter from the under-dense high-energy shower front efficiently. Therefore, in the following, this direct cascade scattering component will be neglected.

\subsection{Uncertain plasma properties}
There are two main uncertainties concerning the plasma, the first uncertainty is the lifetime of the free charges, the second is the free charge collision frequency. In~\cite{dVries15}, we considered plasma lifetimes as measured in~\cite{Ver78,Kun80,Haa83a,Haa83b}, where two different plasma constituents have been measured after irradiating a block of ice with either X-rays or 3~MeV electrons. The first plasma constituent was a relatively short-lived free electron plasma with lifetimes ranging from 0.1~ns for ice at relatively high temperatures around $0^\circ$~Celsius, up to several tens of nanoseconds for temperatures below $-60^\circ$~Celsius. The second plasma constituent was a rather long-lived proton-like plasma with lifetimes ranging from tens of nanoseconds at relatively high temperatures up to $1\;\mathrm{\mu s}$ at temperatures below $-60^\circ$~Celsius. As will become even more prominent in this article, the feasibility of the radar detection technique for detecting neutrino induced particle cascades in the PeV-EeV range, is crucially dependent on the lifetime of the plasma.

In addition to the issue of the lifetime of the plasma, there is also the issue of the free charge collision rate. In case the collision frequency is larger than the detection frequency, the plasma will not be able to scatter the radio wave efficiently. To calculate the collision rate, the so-far unknown (in)-elastic scattering cross-section for a free charge (electron or proton-like) in ice is needed. Since the lifetime of the different plasma constituents in ice has been found to change over several orders of magnitude as function of ice temperature~\cite{Ver78,Kun80,Haa83a,Haa83b}, a similar behavior can be expected for the (in)-elastic scattering cross-section. It follows that the free charge collision frequency for the different plasma components in ice is unknown. In the following we will therefore assume an overall efficiency factor $\eta$ for the radar scattering process. It is clear that additional experimental efforts are required to determine the scattering efficiency and the effect of the free charge collision frequency.

\section{The radar return power}
\seclab{lt}
In the previous section we determined both the under-dense radar scattering cross-section, as well as the over-dense radar scattering cross-section. The total cross-section can thus be decomposed as,
\beq
\sigma=\sigma_{\rm ud}+\sigma_{\rm od}.
\eeq
The radar return power $P_{\rm r}$ for a bi-static radar configuration is now given by,
\beq
P_{\rm r}=P_{\rm t} \frac{\eta G\sigma}{4\pi(R_1)^2}\frac{A_{\rm eff}}{4\pi(R_2)^2}e^{-4|R_1+R_2|/L_{\rm att}},
\eeq
where, $P_{\rm t}$ is the transmit power, $A_{\rm eff}$ is the effective area of the receiver antenna, $R_1=|\vec{x_1}-\vec{x_c}|$ is the distance from the transmitter located at $\vec{x_1}$ to the cascade located at $\vec{x_c}$ and $R_2=|\vec{x_2}-\vec{x_c}|$ is the distance from the cascade to the receiver located at $\vec{x_2}$. We account for the attenuation in the medium by considering the frequency dependent attenuation length $L_{\rm att}(\omega)$. Next to this we also included the scattering efficiency parameter $\eta$ to account for the uncertain plasma parameters as discussed in the previous section. The factors $4\pi$ in Eq.~(23) assume isotropic scattering, which is corrected by the transmitter beaming factor $G$, where the geometry factor $f_{\rm geom}$, which is directly incorporated in the over-dense radar scattering cross-section (Eqs.~(11-14)), covers the non-isotropic scattering off the plasma.

Since the plasma has a finite lifetime, the return signal is limited in time which leads to a dispersion around the transmit frequency. An example of this effect is given in~\figref{dispersion}, where we show the frequency response of a 50~MHz scattered signal for a plasma which lives over different lifetimes.
\begin{figure}
  \centering{
  \includegraphics[height=.3\textwidth]{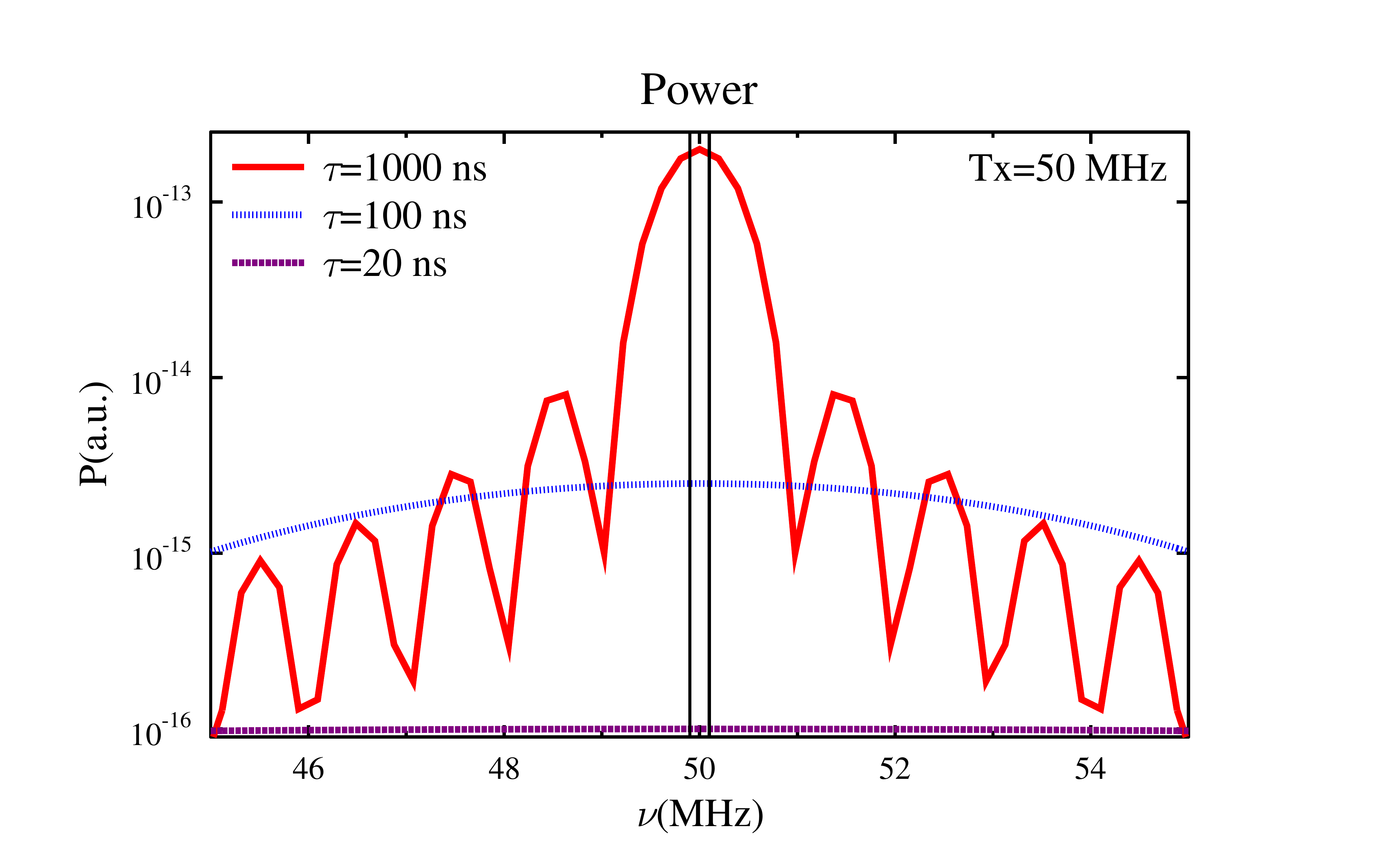}}
\caption{The frequency response for a 50~MHz return signal for different plasma lifetimes. Due to the limited amount of oscillations the plasma is able to re-emit within the lifetime of the plasma, the return signal will be dispersed around the transmit signal. The vertical black lines denote a $\Delta \nu= 100$~KHz detection bandwidth around the transmit frequency of 50~MHz.}
\figlab{dispersion}
\end{figure}

It follows that the signal power will depend on the bandwidth, $\Delta \nu$, in which the signal is detected. In the following, these effects will be taken into account by evaluating the return power over the detection bandwidth $\Delta\nu$,
\beq
P_r(\nu,\Delta \nu,\tau) = \int\limits_{\nu-\Delta \nu}^{\nu+\Delta \nu} P_r(\nu,\tau),
\eeq
where,
\beq
P_r(\nu,\tau) = \int \mathrm{dt}\; e^{2\pi\nu t} P_r(t)\Pi(\tau),
\eeq
is given by the convolution of the return power for infinite lifetime $P_r(t)$, and the rectangular function $\Pi(\tau)$, resulting in a $sinc$ function in frequency space in case $P_r(t)$ oscillates at a fixed frequency.

To determine the sensitivity of the radar detection technique, we first need to determine the noise level. Since we are searching for a signal scattered off the static ionization plasma, the return signal will be emitted at the same frequency as the transmitted signal. For the considered frequency range in this article the main noise will be due to system noise and is given by,
\beq
P_n = k_b T \Delta \nu,
\eeq
where $k_b$ is Bolzmann's constant, and $T=325$~K is the noise temperature which will be taken similar to its value as has been obtained at the South-Pole by the ARA collaboration~\cite{ARA}. Hence, both the signal power, as well as the noise power depend on the detection band-width. Nevertheless, the relative increase in signal power is expected to drop rapidly with increasing detection bandwidth, where the noise scales linearly and it follows that a small detection window around the transmit frequency is favored.

\section{Sensitivity}
As a first application of our improved modeling effort, we will in this section calculate the sensitivity for a simplified detector. To do this, we generated a set of neutrino induced particle cascades for primary neutrino energies of $E=10^{15}-10^{20}$~eV. Due to the Earth absorption effect, at the considered energies no neutrino events are expected to come from below the horizon, hence we only consider neutrinos in a zenith range between $\theta=0^\circ-90^{\circ}$, where $0^\circ$ means the neutrino comes from the zenith and $\theta=90^\circ$ denotes a neutrino coming from the horizon. Furthermore, we require that a significant part of the neutrino energy is deposited in the high-energy cascade. As such we restrict ourselves to (anti-)electron charged current, neutral current and Glashow~\cite{Glashow} induced interactions. We force the neutrinos to interact within a user defined interaction volume which is chosen as a cylinder with a 2~km radius and a height equal to the depth of the South-Pole ice layer, $z=2778$~m. Each event is accordingly given a weight $w\ [s^{-1}]$ corresponding to,
\begin{equation}
 w_i=\frac{dN_{\rm expected}}{dt\ dA\ d\Omega\ dE}\cdot \left(\frac{dN_{\rm simulated}}{dA\ d\Omega\ dE}\right)^{-1}\ .
\end{equation}
Integrated over the full detector volume, these weights thus represent the ratio of the expected rate of events to the number of simulated events.

The (non-optimized) detector configuration is shown in~\figref{grid_radar}. We consider 5 transmit antennas emitting at 1~kW in the vertical (z) and horizontal (x,y) polarizations, leading to a total emitted power of 15~kW. The antennas are positioned 1~km apart, and each antenna is surrounded by 4 receiver antennas which are positioned 500~m to the North (y), East (x), South (-y), and West (-x) of the transmit antennas. In total this gives us 5 transmit antennas and 16 receiver antennas. 
\begin{figure}
  \centering{
  \includegraphics[height=.4\textwidth]{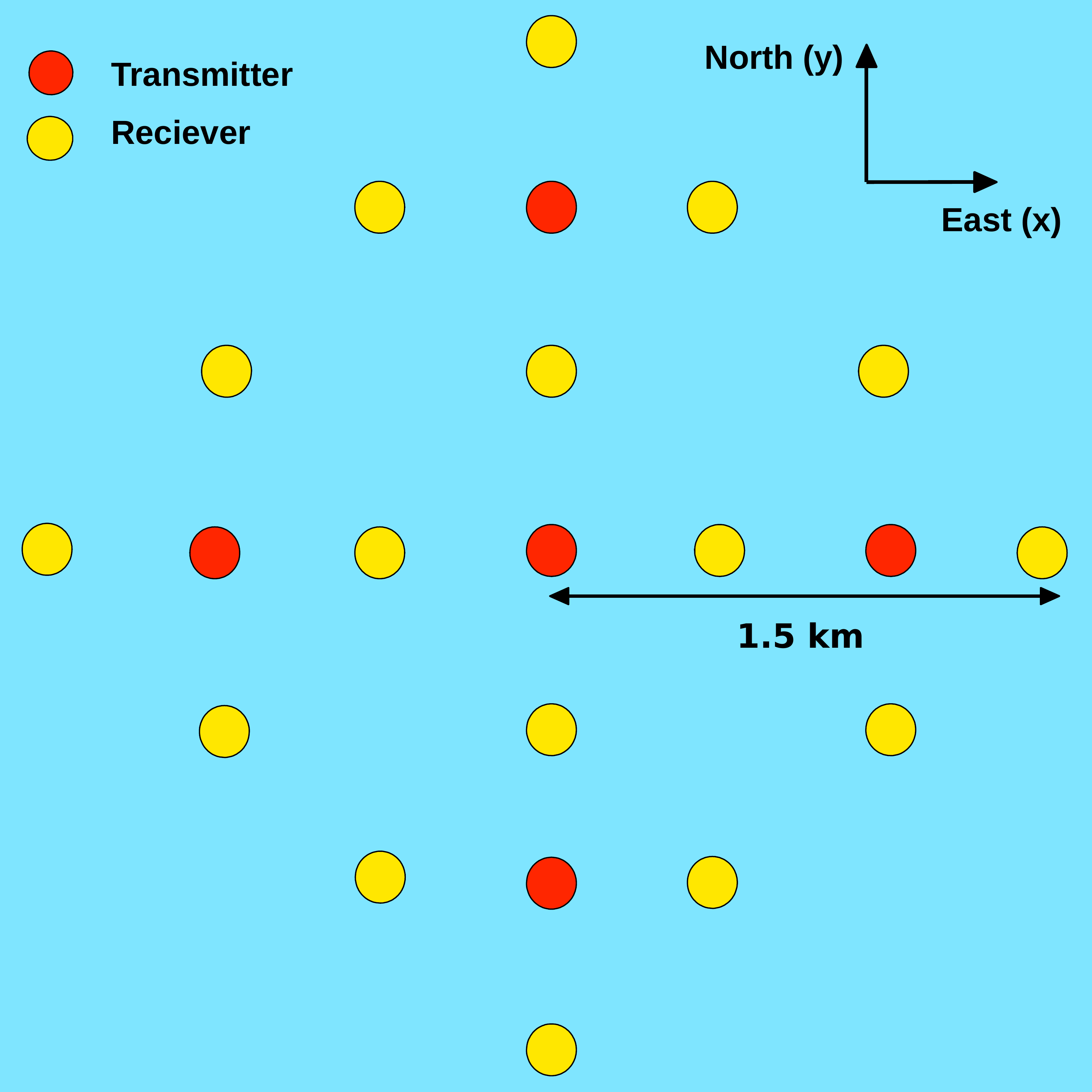}}
\caption{The (non-optimized) detector configuration. The red dots denote the transmitters which are separated 1~km from each other. Each transmit antenna is surrounded by 4 receiver antennas positioned 500~m to the North (y), East (x), South (-y) and West(-x), shown by the yellow dots. Figure adapted from~\cite{dVries17b}.}
\figlab{grid_radar}
\end{figure}
To obtain a sensitivity, we will first determine the effective area of our detector. This will be done considering two types of plasma, a long lived free proton plasma with a lifetime $\tau_p=1\;\mathrm{\mu s}$, which will be detected at 50~MHz, within a detection bandwidth $\Delta \nu=0.1$~MHz, and a shorter lived free electron plasma with a lifetime of $\tau_e=20$~ns which will be probed with a transmit frequency of 450~MHz, within a detection bandwidth of $\Delta \nu = 0.1$~MHz. The lifetimes are taken from~\cite{Ver78,Kun80,Haa83a,Haa83b}, where it should be noted that the obtained lifetimes vary within a range of O(10~ns) at relatively high temperatures up to $1\;\mathrm{\mu s}$ at temperatures around -60 degrees Celsius for the free proton plasma, and from 0.1 ns up to tens of nanoseconds for the free electron plasma again for both high and low ice temperatures. Since the ice temperature at the South-Pole is around -50 degrees Celsius we consider rather progressive values for the lifetime. The attenuation length $L_{att}$ was chosen conservatively following the procedure given in~\cite{dVries15}, which is based on a parameterization from measurements performed at the Ross ice-shelf~\cite{Bar12}.

Having fixed the plasma properties, we can now determine the sensitivity of the detector. The sensitivity is defined following~\cite{ARA}. Assuming no signal is detected, an upper-limit is set on the normalization $\Phi_0$ of the neutrino spectrum,
\begin{equation}
 \Phi(E) = \Phi_0\times E^{-\gamma}\ ,
\end{equation}
here assumed to be an unbroken power-law with spectral index $\gamma$. To constrain $\Phi_0$, we calculate two upper-limits. The first assumes a spectral index $\gamma=2$, which is typically expected for high-energy cosmic neutrino sources. This allows integrating over the full energy range. Using the Feldman-Cousins method~\cite{Fel98} to obtain the upper limit at 90\% significance then results in,
\begin{equation}
 \Phi_0 \leq \frac{2.44}{\Delta t \int{\mathcal{A}_{\rm eff}\frac{d\log(E)}{E}}}\ .
\end{equation}
In this expression, $\mathcal{A}_{\rm eff}$ $[\mathrm{cm^2\ sr}]$ is the effective area of the detector, defined as
\begin{equation}
 \mathcal{A}_{\rm eff}(E) \equiv \frac{1}{\Phi}\cdot \frac{d\dot{N}_d}{dE}\cdot,
\end{equation}
where $\dot{N}_{\rm d}  = \sum_i{w_i\cdot f_i}$ is the rate at which events are detected. Here, the trigger condition $f_i$ is given by $f_i=1$ in case at least one of the receiver antennas observes a signal above the background. In case no antenna is triggered, $f_i=0$. 

Furthermore, we also consider the differential sensitivity, defined as the 90\% upper limit of $\Phi_0$ in case no events are detected over a decade of energy. Using the Feldman-Cousins method the upper limit is obtained by,
\begin{equation}
 \Phi_0 \leq 2.44\cdot \frac{E}{\ln(10)\cdot \Delta t\cdot \mathcal{A}_{\rm eff}}\ ,
\end{equation}
which assumes that $\frac{d\dot{N}_d}{d\ln(E)}$ is a constant over the interval $E\in \left[E_c/\sqrt{10}, \sqrt{10}E_c\right]$, set equal to the value at its center $\frac{d\dot{N}_d}{d\ln(E)}(E_c)$.

\begin{table*}[t]
\small
\begin{tabular}{lcccc}
\hline
				Plasma type & Scattering efficiency & IceCube astrophysical unbroken & Kotera maximal flux & Ahlers ($E_{\rm min} = 10^{17.5}$~eV) \\
\hline
proton    &$\eta$ = 1.		&	29.3	& 24.8 	& 9.8\\
proton    &$\eta$ =0.01		& 	11.1	& 10.5 	& 5.8\\
electron  &$\eta$=1.		& 	2.2 	& 1.8 	& 0.6 \\
electron  &$\eta$ = 0.01	&	0.3 	& 0.2	 & 0.07 \\
\hline
\end{tabular}
\caption{The number of expected neutrinos per year for the detector configuration shown in~\figref{grid_radar} considering different plasma configurations and scattering efficiencies. The expected event number is given considering an extrapolation of the IceCube astrophysical flux given an unbroken power law spectrum, the Kotera maximum flux and the Ahlers flux with $E_{\rm min}=10^{17.5}$~eV.}
\tablab{tab:numneutrinos}
\end{table*}

\figref{Sensitivity} shows the expected sensitivity for the radar detector in one year of data taking for the scattering off the free electron plasma (top), and for the scattering off the free proton plasma (bottom) for two values of the scattering efficiency $\eta$. The radar sensitivity is compared with the astrophysical neutrino flux measurement from IceCube~\cite{HESE4year} and the ARA 37 projected sensitivity for 3 years, as well as the GZK flux predictions taken from~\cite{Ahlers2010} and~\cite{Kotera2010}. 
\begin{figure}[t]
\begin{minipage}{.4\textwidth}
\centering{
  \includegraphics[height=.7\textwidth]{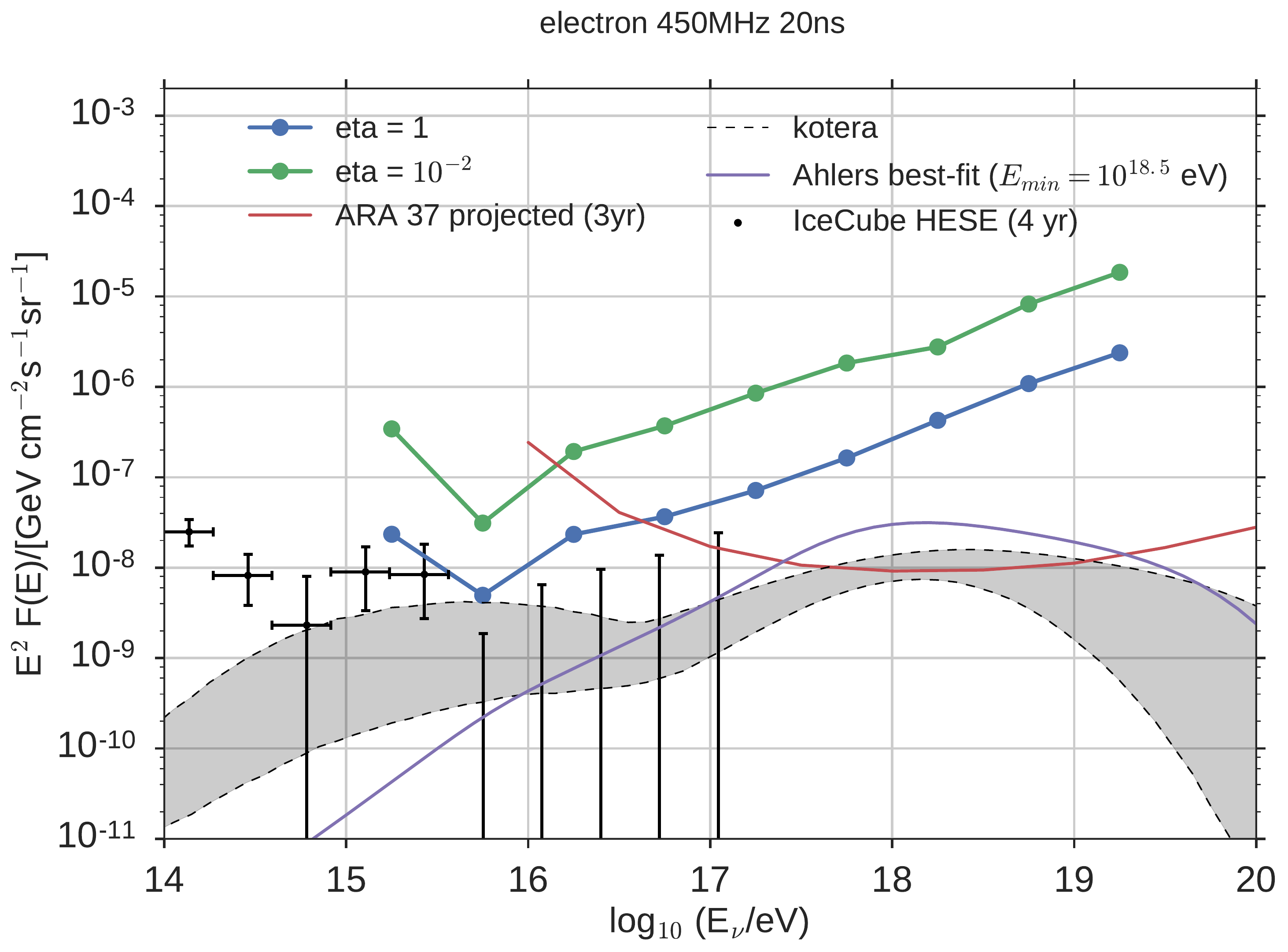} \includegraphics[height=.7\textwidth]{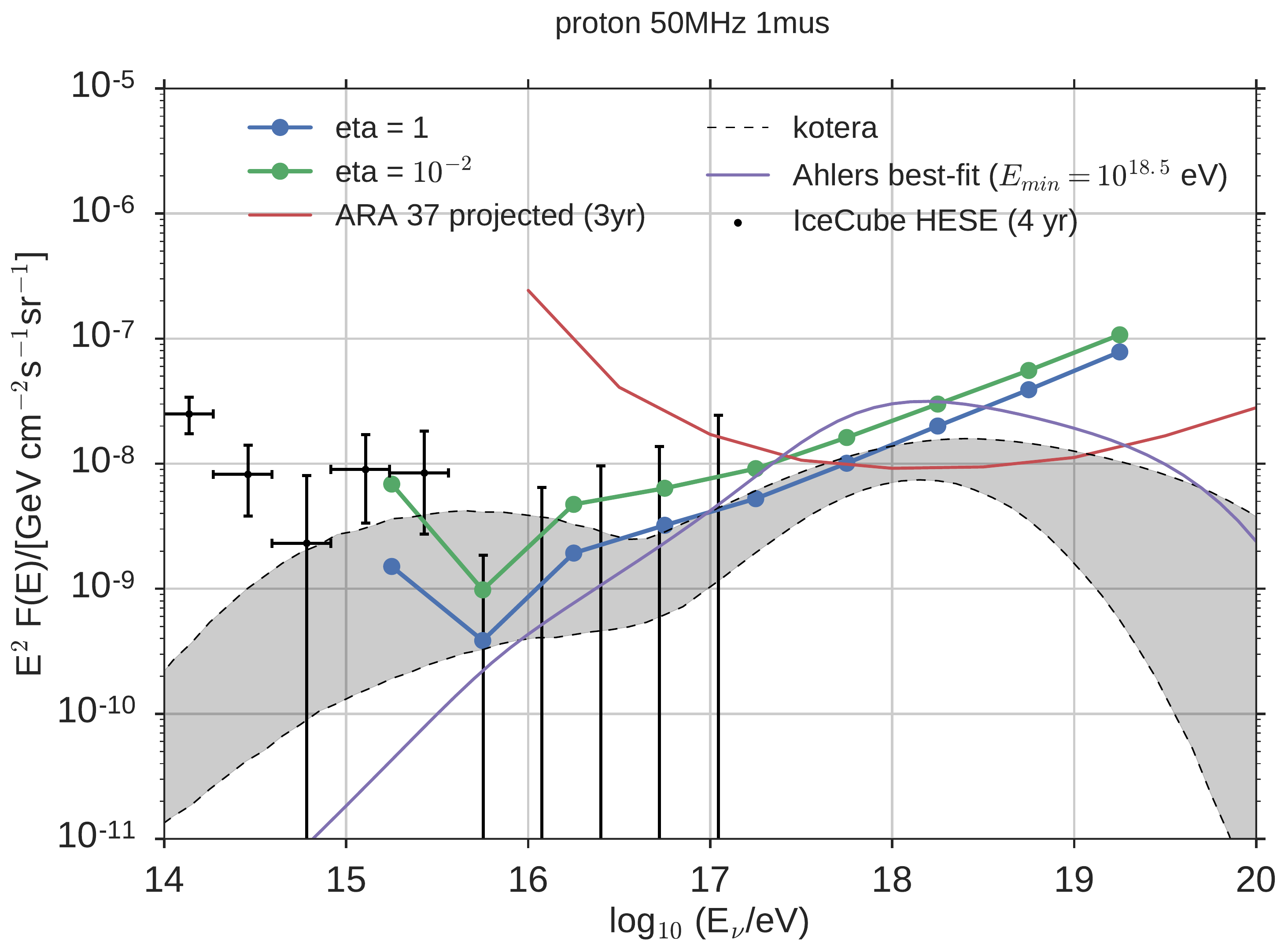}}
\end{minipage}%
\caption{Detector sensitivity for the set-up given in~\figref{grid_radar}: \emph{top)} scattering off a free electron plasma with lifetime $\tau=20\;\mathrm{ns}$, transmit frequency $\nu_t=450\;\mathrm{MHz}$;  \emph{bottom)} scattering off a free proton plasma with lifetime $\tau=1\;\mathrm{\mu s}$, transmit frequency $\nu_t=50\;\mathrm{MHz}$. The transmit power is $P_t=1$~kW. The detection band-width is chosen to be $\Delta \nu=0.1$~MHz. The detector sensitivity is compared with recent measurements of the astrophysical flux, ARA 37 and the most reasonable GZK models taken from~\cite{Ahlers2010} and~\cite{Kotera2010}.}
\figlab{Sensitivity}
\end{figure}
\begin{figure}[t]
  \centering{
  \includegraphics[height=.3\textwidth]{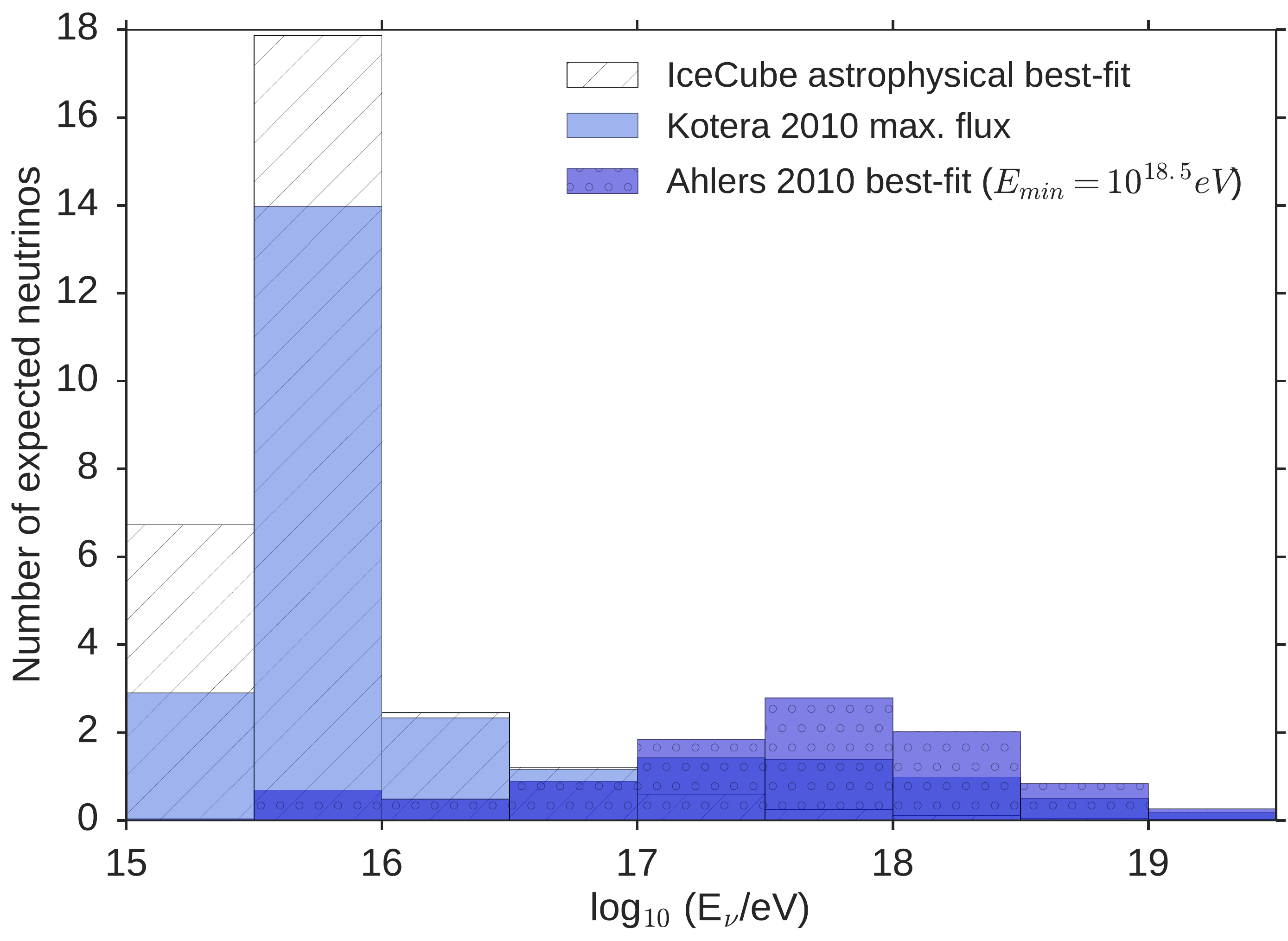}}
\caption{The expected number of events in half-decade energy bins for the most favorable situation of a free proton plasma with a lifetime of 1~$\mu$s, probed at 50~MHz and 100\% scattering efficiency. The number of events is given for the set-up shown in~\figref{grid_radar} for an integration time of one year. Three different neutrino fluxes are considered, an unbroken astrophysical flux as detected by IceCube, the Kotera maximum flux, and the Ahlers flux considering $E_{\rm min}=10^{17.5}$~eV. }
\figlab{proton_eta1}
\end{figure}
Assuming the astrophysical neutrino flux measured by IceCube and the GZK neutrino flux estimation from~\cite{Ahlers2010} and~\cite{Kotera2010}, the number of neutrinos expected to trigger at least one of the antennas in our radar detector are reported in \tabref{tab:numneutrinos}.

\figref{proton_eta1} shows the number of neutrinos versus energy which are expected to be seen with the radar detector layout presented in this paper within one year of operation from different model predictions. For this figure we assume the most favorable situation given by a free proton plasma with a lifetime of 1~$\mu s$, probed at 50~MHz, assuming a scattering efficiency of 100\%.  In addition, the expected number of astrophysical neutrinos assuming an unbroken power law flux ($E^{-2.13}$) as suggested from the IceCube measurement is plotted.
   
It follows that, depending on the scattering efficiency, the radar detection technique should be able to probe both the high-energy tail of the cosmic neutrino flux measured by IceCube, being very sensitive at the Glashow resonance around 6.3~PeV, as well as the GZK neutrino flux. 

\section{Summary and conclusions}
In this article we present a detailed model for the radar scattering off a high-energy neutrino induced particle cascade in ice. In previous work we considered a simplified scattering geometry, as well as an empirically derived scattering cross-section. In this article we extended our model to a general scattering geometry. In addition to this extension, we also present a detailed model of the scattering cross-section. This is done by considering both the under-dense scattering as well as the scatter off the over-dense plasma region. Furthermore, we discuss that the scattering of the high-energy cascade front can be neglected. The main scattering contribution is given by the over-dense scattering region, which is modeled taking into account skin effects, as well as the diffraction pattern of such a scatter. Furthermore, it is shown that for an efficient scatter the lifetime of the plasma has to be relatively large compared to the radar detection frequency. 

As a first application of the derived model, we calculate the effective area and sensitivity for a simplified detector set-up. The set-up consists out of 4 transmit antennas and 16 receivers placed 500 meters apart on a 4.5~km$^2$ surface area. It is shown that the radar detection technique is a very promising method to probe the high-energy cosmic neutrino flux above PeV energies, shifting into the expected GZK neutrino flux in case one is able to scatter efficiently ( $> 1\%$ for a free proton plasma). The scattering efficiency, however, depends on several plasma properties such as the free charge collision frequency, as well as the lifetime of the plasma. These properties are up to now badly known, and further experimental work is thus encouraged.

\section{Acknowledgment}
The authors would like to acknowledge the following funding agencies for their support to the research presented in this work: The Flemish Foundation for Scientific Research (FWO-12L3715N - K.D. de Vries), the FWO Odysseus program (G.0917.09. - N. van Eijndhoven) and the FOM program (12PR304 - O. Scholten). The authors would like to thank S. Prohira for ongoing discussions about the presented work.




\end{document}